# Assessment of Display Performance and Comparative Evaluation of Web Map Libraries for Extensive 3D Geospatial Data

Toshikazu Seto [1,2] *, Yohei Shiwaku [3], Takayuki Miyauchi [3], Daisuke Yoshida [4], Yuichiro Nishimura [5]

[1] Department of Geography, Komazawa University, Tokyo, Japan – tosseto@komazawa-u.ac.jp
[2] Center for Spatial Information Science, the University of Tokyo, Chiba, Japan
[3] Geolonia Inc., Tokyo, Japan – info@geolonia.com
[4] Graduate School of Informatics, Osaka Metropolitan University, Osaka, Japan – daisuke@omu.ac.jp
[5] Faculty of Letters, Nara Women's University, Nara, Japan – nissy_yu@cc.nara-wu.ac.jp

**Abstract**

Large-scale 3D geospatial data visualization has become increasingly critical for the development of the digital society infrastructure in Japan. This study conducted a comprehensive performance evaluation of two major WebGL-based web mapping libraries, CesiumJS and MapLibre GL JS, using large-scale 3D point-cloud data from the VIRTUAL SHIZUOKA and PLATEAU building models. The research employs standardized 3D Tiles 1.1, and Mapbox Vector Tiles (MVT) formats, comparing performance across different data scales (2nd and 3rd grid levels) using Core Web Vitals metrics, including First Contentful Paint (FCP), Largest Contentful Paint (LCP), Speed Index, Total Blocking Time (TBT), and Cumulative Layout Shift (CLS).
The results demonstrate that MVT-based building visualization with MapLibre GL JS achieves optimal performance (FCP 0.8s, TBT 0ms), whereas MapLibre GL JS combined with deck.gl shows superior performance for large-scale point cloud processing (TBT: 3ms, CesiumJS: 21,357ms). This study provides data-driven selection guidelines for appropriate technology choices according to use cases, establishing reproducible performance evaluation frameworks for 3D web mapping technologies during the WebGPU and OGC 3D Tiles 1.1 standardization era.

## 1. Introduction

Japan has recently witnessed rapid advancements in large-scale 3D urban model development through initiatives such as the Ministry of Land, Infrastructure, Transport, and Tourism in Japan (MLIT) Project PLATEA. Project PLATEAU has completed 3D model data acquisition for over 250 local governments (approximately 30,000 km²) with more than 23 million CityGML format data entries as of 2025 and plans to expand to 500 cities by 2027. Meanwhile, VIRTUAL SHIZUOKA in Shizuoka prefecture has developed a 30TB point cloud dataset covering the entire Shizuoka Prefecture (6,700 km²), providing an open digital twin at a 1:1 scale. Efficient web-based visualization of these large-scale datasets presents a critical technical challenge for building Japan's digital society infrastructure.

This study focused on Numazu City, Shizuoka Prefecture, where both datasets were developed comprehensively. This research evaluates the display performance using building models (CityGML format) from PLATEAU and point cloud data (LAS format) from VIRTUAL SHIZUOKA. The data conversion process utilizes the 3D Tiles 1.1 specification, approved as an OGC standard in 2023, and the Mapbox Vector Tiles (MVT) format. 3D Tiles 1.1 represents a significant evolution from version 1.0, incorporating an optimized hierarchical level of detail (HLOD), implicit tiling functionality, multigranularity semantic metadata support, and direct integration of glTF 2.0, substantially improving the streaming delivery performance for large-scale geospatial data.

The display performance evaluation compared two major WebGL-based web mapping libraries: CesiumJS and MapLibre GL JS (integrated with deck.gl and loaders.gl). CesiumJS serves as the de facto standard for Earth-scale 3D visualization, excelling in the efficient streaming of large datasets and HLOD management through optimized rendering pipelines that natively support 3D tiles. In contrast, MapLibre GL JS features a vector tile-specific architecture that provides high-performance 3D rendering capabilities through deck.gl integration. Both libraries employ distinct design philosophies and optimization strategies, requiring detailed performance characteristic comparisons according to the intended use cases.

The objective of this study is to conduct a quantitative comparative analysis of CesiumJS and MapLibre GL JS across different data formats (3D Tiles and MVT) and different scales (10km grid square and 1km grid square levels), establishing clear selection guidelines for optimal technology choices according to application requirements.

## 2. Research and Technical Background

### 2.1 Evolution of 3D Web Mapping Technology

The 3D web mapping technology has developed rapidly with the adoption of WebGL. The MLIT conducted a comprehensive investigation in the "Survey Report on WebGIS Development Optimized for 3D Urban Models" (MLIT, 2024), examining possibilities beyond CesiumJS. This investigation revealed that WebGPU based architecture are promising web mapping engines that can dynamically change styles, render aesthetically sophisticated graphics, and use advanced expression techniques.

PLATEAU VIEW currently uses CesiumJS as a WebGL-based 3D mapping engine suitable for rendering 3D urban models and terrain. However, compared to other mapping engines, they lack various data-visualization capabilities. Performance issues arise from the high computational demands required for precision rendering, which can potentially degrade application performance. In addition, CesiumJS performs rendering processing internally within the engine, making direct rendering customization by application developers challenging. The standardisation of data formats represents the next critical step in refining and enhancing mapping applications, building on foundational developments in 3D web mapping technology.

---

* Corresponding author

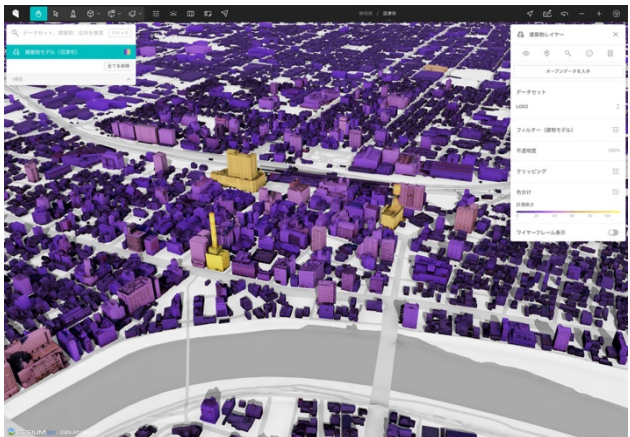

Figure 1. PLATEAU view interface.
(Building colors indicate height levels in stages)

**2.2 Data Format Standardization**

The 3D Tiles 1.1 specification, approved as an OGC standard in 2023, represents a significant evolution from version 1.0. The previous 3D Tiles 1.0 employed proprietary formats with custom headers appended to glTF, whereas version 1.1 enabled tiles to be described as pure glTF files. Metadata are expressed as glTF extensions, enhancing interoperability with existing glTF ecosystems.

Mapbox Vector Tiles (MVT) represent the vector tile specifications established by Mapbox and packaging vector data using Protocol Buffers. The latest version is 2.1 (2016), which is optimized for the web delivery of 2D data such as building footprints. According to Netek et al. (2020), the MapLibre Tile (MLT) format research demonstrates Pareto improvements in both the compression ratio and decode speed for vector tiles, with large-scale tilesets no longer requiring complex vector tile schema-specific optimization.

**2.3 Prior Research on Performance Evaluation Methods**

Zunino et al. (2020) conducted the most comprehensive web mapping library performance evaluation. This study evaluated three major libraries, Leaflet, Mapbox GL JS, and OpenLayers, implementing both raster and vector tiles. The evaluation employed automated performance testing via Puppeteer and implemented detailed performance measurements using the Chrome DevTools Performance tab. Testing encompassed 14 different device types (smartphones, tablets, and PCs) under standardized conditions of 10Mbit/s WiFi connectivity across six distinct geographic locations with automated pan and zoom operations. The statistical analysis included ten repetitive test runs for each configuration.

Key findings showed that Mapbox GL JS achieved superior overall performance on mid-to-high-end devices, with OpenLayers demonstrating the best performance across all device categories for raster tiles. Vector tile implementations showed clear network efficiency advantages (a 33% reduction in bandwidth usage). For low-performance devices, raster tiles proved advantageous owing to CPU/GPU limitations, whereas Leaflet's vector tile implementation showed a significantly lower performance owing to third-party plugin constraints. This study utilized real-world Argentine census data (52,408 census areas), representing the significant characteristics of practical evaluation.

Netek et al. (2019) conducted performance evaluation focused on large-scale dataset processing in "Performance Testing on Marker Clustering and Heatmap Visualization Techniques: A Comparative Study on JavaScript Mapping Libraries." This study compared five libraries—Leaflet.markercluster, Supercluster, Mapbox GL JS, PruneCluster, and OpenLayers— by evaluating the marker clustering and heatmap visualization performance. The evaluation employed the Google Chrome DevTools Performance tab for detailed measurements, implementing systematic testing from 10,000 to 3,000,000 point datasets. Standardized testing under Intel Core i5-6200U and 8GB RAM conditions utilized the nature conservation database as a real data source. The evaluation metrics included the loading time (milliseconds), rendering performance across data quantities, memory usage patterns, visualization quality assessment, and performance degradation threshold identification. The principal findings showed that Mapbox GL JS achieved superior overall performance through WebGL technology, with 100,000 points identified as a critical performance threshold for most libraries. Datasets exceeding 500,000 points rendered some libraries completely non-functional, with vector-based rendering approaches demonstrating scalability advantages. Large-scale dataset visualization requires WebGL acceleration.

Netek et al. (2020) implemented a detailed performance comparison of vector and raster tiles. This study evaluated technical characteristics using the Czech Republic OpenMapTiles (OMT) project data, municipal boundaries, and OpenStreetMap airport layers. The evaluation employed semi-automated testing with consistent pan-zoom operations, multibrowser testing across four browsers, server load pattern analysis, and network traffic monitoring. The evaluation metrics included Time to Interactive (TTI) measurement, First Contentful Paint (FCP) analysis, network transfer measurement, and loading performance indicators. Technical comparison results showed that pre-generated raster tiles achieving a 400ms faster map loading time than vector tiles, whereas vector tiles demonstrated a 20-50% smaller download size. Data size analysis revealed that WebP raster tiles require one-third of the data volume of PNG raster tiles, whereas vector tiles require substantially fewer server requests. High-speed connections favored vector tiles, whereas low-speed connections favored raster tiles. This research recommends hybrid approaches that utilize vector tiles for interactive layers and raster tiles for base maps. The advantages of vector tiles include small file sizes, fewer server requests, and dynamic styling capabilities, whereas those of raster tiles include rapid initial loading and superior performance for low-speed devices and connections.

Fujimura et al. (2019) presented "Design and Development of the UN Vector Tile Toolkit" at FOSS4G 2019, reporting on UN implementation of vector tile technology. This study represents an important practical adoption of vector tiles by public institutions. The UN Vector Tile Toolkit utilizes existing open-source software, including Tippecanoe, Maputnik, and Vector Tile optimizer, implemented as Node. j-based scripts. This research addressed four challenges: continuous and automated global base map vector tile updates, mobile field application usability, and interoperability with existing enterprise geospatial frameworks. The performance evaluation achieved successful processing of global OpenStreetMap data (45GB) in approximately 80 h (3, 7, and 22 min), representing a substantial reduction in processing time compared to conventional OpenMapTiles, which require 37 d (approximately 900 h). This study provides important practical performance metrics for large-scale geospatial data vectorization.

## 3. Study Area and Regional Characteristics

Numazu City of Shizuoka Prefecture was selected as the study area. Numazu City possesses extremely diverse and complex regional characteristics that are ideal for 3D urban modelling case studies. As one of the cities selected under Project PLATEAU, three core characteristics—topographic diversity, appropriate urban scale, and composite industrial structure—position it as a priority subject for 3D geospatial data investigations.

Numazu City occupies the eastern Shizuoka Prefecture at the base of the Izu Peninsula, with an area of 186.84 km² and a population of approximately 180,000, representing a mid-scale regional city. Located at 138°52'E, 35°06'N, with 64.491 km of coastline facing Suruga Bay, this location provides a geographically favourable environment surrounded by Japan's representative geographical features, including Mount Fuji, the Hakone mountain range, and Suruga Bay. An urban population density of approximately 2,033 people/km² maintains a moderate density, forming a research-appropriate urban scale that is neither excessively dense nor sparse.

Numazu City's most noteworthy characteristic is the extremely large elevation differential from sea level to 1,333m, and the corresponding diverse geomorphological elements. The composition of the volcanic terrain featuring Atagoyama as the highest peak, the alluvial plain formed by the Kano River, and Japan's deepest Suruga Bay (maximum depth 2,500m) provides incomparable diversity for 3D modelling technology verification.

Numazu City's decisive advantage for 3D geospatial data research stems from the advanced data infrastructure development under Project PLATEAU. Selected as one of first 56 cities with continuous 3D urban model data development and updates since 2020, comprehensive CityGML 2.0-compliant 3D urban models were established at LOD 0-2 levels, achieving coverage of 187.10 km² for LOD1 and 4.23 km² for LOD2. Data are provided in WebGIS-compatible formats, including 3D Tiles and MVT, and distributed as open data through the G-Spatial Information Center.

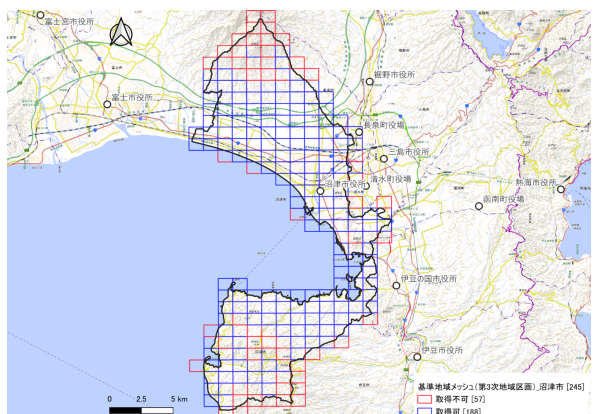

Figure 2. Study area of Numazu city (1km grid).

## 4. Data Conversion and Web Map Display Performance Evaluation

### 4.1 Target Data and Conversion Process

**4.1.1 Source Data Overview**: This study utilized the following data for Numazu City, Shizuoka Prefecture: 3D point cloud data (LAS format) sourced from Shizuoka VIRTUAL SHIZUOKA are provided at the fundamental map information level of 500 units (vertical 300m × horizontal 400m) according to the national map sheet divisions. The total file size for the entire city coverage is approximately 462GB (LAS format). The project PLATEAU building model 3D urban models (CityGML format) were provided in 3rd mesh (1 km grid square) units. The total file size for the entire city coverage was approximately 2.3GB (CityGML format). The evaluation target ranges encompassed two scales: the 2nd mesh (523856: 10km grid square) and the 3rd mesh (52385628: 1km grid square), enabling a systematic analysis of the effects of data quantity on display performance.

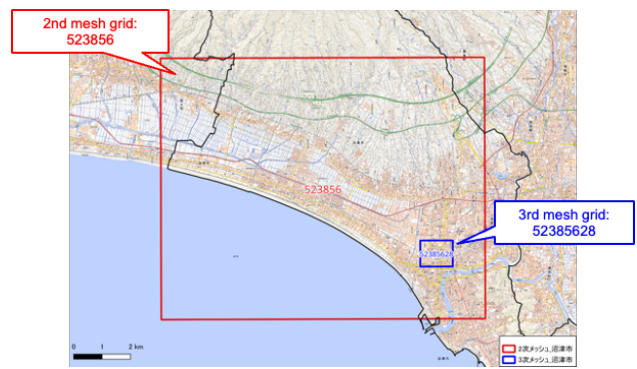

Figure 3. Spatial definition of 2rd and 3rd grid in Numazu.

**Data Conversion Procedures**: Point cloud data (from LAS to 3D Tiles) conversion employed PDAL v2.8.1 (included with QGIS 3.34.12) and py3dtiles v9.0.0. The processing workflow was as follows:

1. LAS file merging via PDAL
2. LAS LAZ format conversion via PDAL
3. coordinate reference system EPSG:6676 (8th system planar rectangular coordinates) assignment
4. X/Y coordinate swapping processing for planar rectangular coordinate system compatibility
5. 3D tile generation via py3dtiles. The 3D tile output from py3dtiles corresponds to version 1.0.

The PLATEAU GIS Converter was used for building model (from CityGML to 3D Tiles) conversion. The output 3D Tiles correspond to version 1.1, with tileset.json recording asset.version as "1.1". The building model (from CityGML to MVT) conversion similarly utilized the PLATEAU GIS Converter with an appropriate zoom-level configuration to produce lightweight 2D representation data.

### 4.2 Development of 3D Web Mapping Display Tools

Five 3D web mapping display tools were developed using the converted data. CesiumJS for point cloud data (3D Tiles 1.0) display, CesiumJS for building models (3D Tiles 1.1) display, MapLibre GL JS + deck.gl + loaders.gl for point cloud data (3D Tiles 1.0) display, MapLibre GL JS + deck.gl for building models (3D Tiles 1.1) display, and MapLibre GL JS for building models (MVT) display comprised of five patterns. Each tool maintained unified perspective settings (near Numazu Station) and background mapping (Geospatial Information Authority of Japan Standard Map: GSIMap), creating comparable environments for the user.

The independently developed dual-screen comparison viewer enables the simultaneous comparison of different datasets and visualization methods. The system under discussion provides parallel display functionality for multiple 3D datasets (point cloud data, building models, and terrain models) within identical geographic ranges. In addition, it integrates synchronised perspectives, unified rendering configuration, and real-time performance metric monitoring.

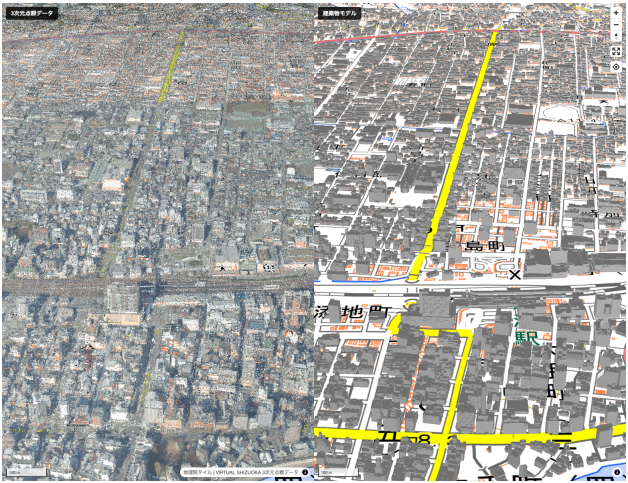

Figure 4. Comparison website for constructed 3D data geographic visualizations.

### 4.3 Display Performance Evaluation Methods

The performance measurement employed the Google Chrome Lighthouse mode, conducting a quantitative evaluation using five indicators based on Core Web Vitals. The evaluation indicators comprised the following: First Contentful Paint (FCP), which measures the time until the initial page content is displayed; Largest Contentful Paint (LCP), which measures the time until the largest element is displayed; Speed Index, which measures the visual progress of page loading; Total Blocking Time (TBT), which measures the cumulative time pages become unresponsive, and Cumulative Layout Shift (CLS), which measures the visual stability of the page layout.

Consideration included the March 2024 Interaction to Next Paint (INP) introduction, which enabled precise measurement of user interaction responsiveness in 3D applications. The evaluation scope encompassed 2nd mesh (523856) and 3rd mesh (52385628) scales, enabling a systematic analysis of the effects of data quantity on display performance.

The measurement protocol enabled cache disabling to exclude cache effects, implementing three measurements per five patterns × two types, and calculating the averages across all three runs.

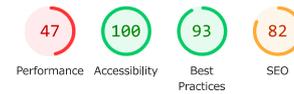
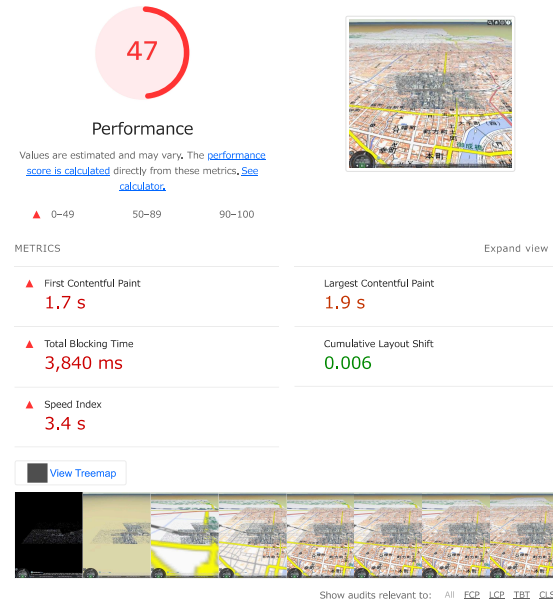

Figure 5. Example of Google Lighthouse evaluation results.

### 5. Results

The display performance evaluation revealed pronounced performance differences according to data format and library combinations (Table1). The best performance was demonstrated by MapLibre GL JS utilizing the MVT format for building display, achieving FCP 0.8 s and Speed Index 0.8 s with TBT 0 ms across both the 3rd and 2nd meshes, demonstrating the advantages of lightweight vector data.

Large-scale point-cloud processing revealed significant differences in the library performance. CesiumJS experienced extreme response delays, with a TBT of 7,270 milliseconds in 3rd mesh and 21,357 milliseconds in 2nd mesh. Conversely, MapLibre GL JS maintained a TBT below 3 ms under identical conditions, confirming the optimization effectiveness of deck.gl and loaders.gl integration. CesiumJS point cloud display recorded FCP 1.6 s, LCP 1.8 s, Speed Index 3.3 s, and CLS 0.006 for the 3rd mesh. 2nd mesh results of FCP 1.6 s, LCP 1.9 s, Speed Index 3.5 s, and CLS 0.005 demonstrated pronounced performance degradation accompanied by an increased data volume. MapLibre GL JS + deck.gl point cloud display achieved FCP 1.4 s, LCP 1.4 s, Speed Index 1.4 s, and CLS 0.001 for the 3rd mesh. 2nd mesh results of FCP 1.4 s, LCP 1.5 s, Speed Index 1.7 s, and CLS 0.001 exhibited stable performance resistant to data scale effects.

The display performance of the building model (3D Tiles) demonstrated that MapLibre GL JS achieved an FCP of 1.3 s compared to CesiumJS 1.6 s, showing certain differences, although the TBT reversal phenomenon with MapLibre GL JS was approximately 1,000 ms versus CesiumJS tens of milliseconds. This indicates that MapLibre GL JS achieves rapid initial loading but experiences main thread-blocking during 3D tile processing.

The CesiumJS building model displayed a recorded FCP of 1.6 s, LCP of 1.8 s, TBT of 3 ms, and CLS of 0.006 for the 3rd mesh. 2nd mesh results of FCP 1.6 s, LCP 2.0 s, TBT 63 ms, and CLS 0.005 demonstrated stable performance. The MapLibre GL JS + deck.gl building model display achieved FCP 1.3 s, LCP 1.3 s, TBT 1,000 ms, and CLS 0.001 for the 3rd mesh. 2nd mesh results of FCP 1.3 s, LCP 1.3 s, TBT 963 ms, and CLS 0.001 were characterized by elevated TBT values. The MVT format-building display exhibited a superior performance across both scales. 3rd and 2nd mesh results of FCP 0.8 s, LCP approximately 1.4 s, Speed Index 0.8 s, TBT 0 ms, and CLS 0.006 demonstrated the advantages of lightweight data. The Cumulative Layout Shift (CLS) results demonstrated that MapLibre GL JS achieved extremely stable values of 0.001, whereas CesiumJS ranged 0.005-0.006, quantifying the effects of 3D rendering architecture differences on visualization quality.

## 6. Discussion

### 6.1 Technical Implications

The evaluation results clarified the use-case-specific optimal selection guidelines for MVT for lightweight data, MapLibre GL JS and deck.gl for large-scale point clouds, and CesiumJS for high-precision 3D visualization. The MVT and MapLibre GL JS combination is optimal for simplified 3D building representation, operating extremely lightly while including a standard heatmap and clustering functionality.

Large-scale point cloud processing demonstrated MapLibre GL JS + deck.gl + loaders.gl, achieving superior responsiveness with large-volume point data and substantial TBT improvements, thereby elevating user experience. High-precision 3D visualization confirmed the suitability of CesiumJS for detailed 3D urban model expression, with reconfirmed advantages in handling building attribute information.

The technical significance of this research stems from the comprehensive performance evaluation utilizing practical-scale datasets during the technological innovation period marked by WebGPU proliferation (from 2024 onwards, achieving a maximum 1,000% performance improvement) and OGC 3D Tiles 1.1 specification standardization. This study provides technical insights addressing large-scale datasets generated by Japan's 3D geospatial data infrastructure projects and clarifies the implementation guidelines for web-based 3D GIS applications.

### 6.2 Implications for Future Perspectives

The evaluation framework construction considered compatibility with next-generation technologies, including CityGML 3.0 specification development and WebXR integration, contributing to both the academic advancement and practical deployment of 3D web mapping technology. The developed dual-screen comparison viewer system enables intuitive and quantitative assessment of the performance differences between data formats and visualization libraries.

This research demonstrates academic significance by presenting evaluation criteria that enable diverse options from traditional CesiumJS-only choices according to intended use. During the technological innovation period marked by WebGPU proliferation and 3D Tiles 1.1 standardization, practical technical insights applicable to Japan's large-scale 3D geospatial data infrastructure projects were provided.

With the anticipated improved 3D tile processing capabilities in mapping engines beyond CesiumJS, the performance differences identified in this study serve as important improvement indicators for future technology development. The development of new mapping engines targeting existing engine problem resolution represents long-term possibilities, with the evaluation frameworks established through this research applicable as evaluation foundations for such emerging technologies.

## 7. Conclusions

This study conducted a comprehensive performance evaluation of CesiumJS and MapLibre GL JS by using practical-scale 3D geospatial datasets. The principal achievements include the objective quantification of performance differences through five lighthouse-based indicators, presentation of appropriate technology selection guidelines according to data characteristics and display requirements, and establishment of reproducible and extensible performance evaluation methodologies.

The evaluation results demonstrated that MapLibre GL JS utilizing the MVT format building display achieved superior performance, recording an FCP of 0.8 s, Speed Index of 0.8 s, and TBT of 0 ms across both the 3rd and 2nd meshes, with lightweight vector data advantages definitively demonstrated. Large-scale point cloud processing revealed that MapLibre GL JS + deck.gl achieved pronounced advantages over CesiumJS, with TBT differences of over 20,000 milliseconds TBT differences confirmed.

| No. | Libraries of WebGIS | Evaluation data | Scope of spatial range | Average of all 3 sessions | | | | |
|---|---|---|---|---|---|---|---|---|
| | | | | First Contentful Paint (FCP) | Largest Contentful Paint (LCP) | Total Blocking Time (TBT) | Cumulative Layout Shift (CLS) | Speed Index |
| 1 | CesiumJS | 3d point-clouds (3D Tiles) | 3rd grid-level (52385628) | 1.6s | 1.8s | 7,270ms | 0.006 | 3.3s |
| 2 | CesiumJS | 3d city models (3D Tiles) | 3rd grid-level (52385628) | 1.6s | 1.8s | 3ms | 0.006 | 2.7s |
| 3 | MapLibre GL JS + deck.gl | 3d point-clouds (3D Tiles) | 3rd grid-level (52385628) | 1.4s | 1.4s | 3ms | 0.001 | 1.4s |
| 4 | MapLibre GL JS + deck.gl | 3d city models (3D Tiles) | 3rd grid-level (52385628) | 1.3s | 1.3s | 1,000ms | 0.001 | 1.3s |
| 5 | MapLibre GL JS + deck.gl | 3d city models (MVT) | 3rd grid-level (52385628) | 0.8s | 1.4s | 0ms | 0.006 | 0.8s |
| 6 | CesiumJS | 3d point-clouds (3D Tiles) | 2nd grid-level (523856) | 1.6s | 1.9s | 21,357ms | 0.005 | 3.5s |
| 7 | CesiumJS | 3d city models (3D Tiles) | 2nd grid-level (523856) | 1.6s | 2.0s | 63ms | 0.005 | 2.1s |
| 8 | MapLibre GL JS + deck.gl | 3d point-clouds (3D Tiles) | 2nd grid-level (523856) | 1.4s | 1.5s | 3ms | 0.001 | 1.7s |
| 9 | MapLibre GL JS + deck.gl | 3d city models (3D Tiles) | 2nd grid-level (523856) | 1.3s | 1.3s | 963ms | 0.001 | 1.3s |
| 10 | MapLibre GL JS + deck.gl | 3d city models (MVT) | 2nd grid-level (523856) | 0.8s | 1.3s | 0ms | 0.006 | 0.8s |

Table1. The display performance evaluation results.

The research contributions extend across academic advancement and practical deployment of 3D web mapping technology. This study holds practical value in supporting critical technology choices that are essential for digital society infrastructure development. The continued development of evaluation frameworks established through this research is anticipated to be a new benchmark criterion during WebGPU proliferation.

## Acknowledgements

The authors acknowledge the support of the project partners. This study was supported by JSPS KAKENHI (grant numbers 22K18505, 23K22032, and 24K15662).

## References

Fujimura, H., Martin Sanchez, O., Gonzalez Ferreiro, D., Kayama, Y., Hayashi, H., Iwasaki, N., Mugambi, F., Obukhov, T., Motojima, Y., Sato, T. 2019: Design and development of the UN Vector Tile Toolkit. *Int. Arch. Photogramm. Remote Sens. Spatial Inf. Sci.*, XLII-4/W14, 57-62. doi.org/10.5194/isprs-archives-XLII-4-W14-57-2019.

Ministry of Land, Infrastructure, Transport, and Tourism (MLIT). 2024. Survey Report on WebGIS Development Optimized for 3D Urban Models. https://www.mlit.go.jp/plateau/file/libraries/doc/plateau_tech_doc_0060_ver01.pdf

Netek, R., Brus, J., Tomecka, O. 2019: Performance Testing on Marker Clustering and Heatmap Visualization Techniques: A Comparative Study on JavaScript Mapping Libraries. *ISPRS International Journal of Geo-Information*, 8(8), 348. doi.org/10.3390/ijgi8080348.

Netek, R., Masopust, J., Pavlicek, F., Pechanec, V.: 2020. Performance testing on vector vs. raster map tiles—comparative study on load metrics. *ISPRS International Journal of Geo-Information*, 9(2), 101. doi.org/10.3390/ijgi9020101.

Open Geospatial Consortium 2023. 3D Tiles Specification 1.1. OGC Community Standard. https://www.ogc.org/standards/3dtiles (20 November 2025).

Tremmel, M., Zink, R.: 2025. MapLibre Tile: A new vector tile format optimized for fast rendering. https://arxiv.org/abs/2508.10791 (20 November 2025).

Zunino, A., Velázquez, G., Celemín, J. P., Mateos, C., Hirsch, M., Rodriguez, J. M.: 2020. Evaluating the Performance of Three Popular Web Mapping Libraries: A Case Study Using Argentina's Life Quality Index. *ISPRS International Journal of Geo-Information*, 9(10), 563. doi.org/10.3390/ijgi9100563.

## Appendix

The source code for the data processing script and comparing viewer is available at GitHub repository.
https://github.com/tossetolab/3dmapcompare/